\documentclass[10]{article}

\usepackage{amssymb}
\usepackage{graphicx}
\usepackage{mathrsfs}
\usepackage{wasysym}
\usepackage{setspace}
\usepackage{lineno}

\begin{document}

\title{The impact of climate change on the structure of Pleistocene mammoth steppe food webs} 




\maketitle

{\center Justin D. Yeakel${}^{1}$, Paulo R. Guimar\~aes Jr${}^2$, Herv\'e Bocherens${}^3$, and Paul L. Koch${}^4$ \\
 ${}^1$Department of Ecology and Evolutionary Biology,}
{\flushleft 
University of California, Santa Cruz \\
1156 High St., Santa Cruz, CA 95064, USA \\
${}^2$Instituto de Ecologia, \\
Universidade de S\~ao Paulo, 05508-900, \\
S\~ao Paulo, SP, Brazil\\
${}^3$Fachbereich Geowissenschaften, Biogeologie \\
Eberhard Karls Universit\"at T\"ubingen \\
H\"oderlinstrasse 12, D-72074 T\"ubingen, Germany \\
${}^4$Department of Earth and Planetary Sciences,\\
University of California, Santa Cruz,\\
1156 High St., Santa Cruz, CA 95064, USA\\
\flushleft{\rm
{\bf Author for Correspondence:} \\
Justin D. Yeakel\\
e-mail: jdyeakel@gmail.com
}

\begin{abstract}

Species interactions form food webs, impacting community structure and, potentially, ecological dynamics.
It is likely that global climatic perturbations that occur over long periods of time have a significant influence on species interaction patterns.
Here we integrate stable isotope analysis and network theory to reconstruct patterns of trophic interactions for six independent mammalian communities that inhabited mammoth steppe environments spanning western Europe to eastern Alaska (Beringia) during the late Pleistocene.
We use a Bayesian mixing model to quantify the contribution of prey to the diets of local predators, and assess how the structure of trophic interactions changed across space and the Last Glacial Maximum (LGM), a global climatic event that severely impacted mammoth steppe communities.
We find that large felids had diets that were more constrained than co-occurring predators, and largely influenced by an increase in {\it Rangifer} abundance after the LGM.
Moreover, the structural organization of Beringian and European communities strongly differed:
compared to Europe, species interactions in Beringian communities before - and possibly after - the LGM were highly modular.
We suggest that this difference in modularity may have been driven by the geographic insularity of Beringian communities.
\end{abstract}


\section{Introduction}
\label{sec:intro}

The structural patterns of species interactions may affect ecosystem dynamics \cite{Allesina:2008vc}, and are sensitive to external perturbations such as climate change \cite{Sole:2010ki,Tylianakis:2010fe}.
Impacts of climate change and other perturbations on food web structure may be immediate or lagged \cite{Tilman:1994p2366}; they can affect communities by
reorganizing interactions \cite{Paine:1980wh},
changing the magnitudes of interactions \cite{KOKKORIS:2002tf,Bascompte:2005vt}, or
eliminating species \cite{Tilman:1994p2366,Sole:2001uu}.
However, observations of community organization across a perturbation event are typically confined to short timescales and to populations with fast turn-over rates.
To assess the long-term effects of climate change empirically, it is necessary to use paleontological or historical information \cite{Dawson:2011bh}.
A climatic perturbation of global significance occurred in the late Pleistocene and culminated with the Last Glacial Maximum (LGM, 26.5 to 19 kyr BP) \cite{Clark:2009be}, strongly impacting mammalian communities worldwide, including the one that extended across the Eurasian mammoth steppe \cite{Yurtsev:2001p3365,FoxDobbs:2008tq}, an environment with no modern analogue \cite{Guthrie:2001vj}.
An examination of species interactions across this climatic event is well-suited to assess the effects of large perturbations on the organization of animal communities.

Although evidence of many paleontological species interactions is irrecoverable, interactions that involve a flow of biomass are recorded in animal tissue and can be reconstructed using stable isotope ratios \cite{Koch:2007ug,Newsome:2007tz,Newsome:WhhVfocb,Yeakel:OfhI8s6v}. 
As a consequence, they can be used to compare patterns of interaction across the mammoth steppe environment.
Mammoth steppe communities were taxonomically similar across Eurasia \cite{Bocherens:2003wo}, although the inherent plasticity of species' roles from Beringia (a region that includes Siberia, Alaska, and the Yukon) to Europe is not known.
Nor is it known whether generalized features of trophic systems, such as the degree of dietary specialization among consumers, varied across this expansive ecosystem.

Global ice sheets attained their maximum volume during the LGM \cite{Clark:2009be}, separating warmer, mesic periods before and after.
This change in global climate had a tremendous impact on the mammoth steppe ecosystem, eliminating temperate species (particularly in Europe), and initiating a shift from tree-covered habitats to xeric grassland-dominated habitats across Eurasia  \cite{Agusti:2002ua}.
Although the mammoth steppe experienced dramatic climatic shifts during the late Pleistocene, whether such changes impacted trophic interactions or by extension community organization, is unknown.
Quantification of trophic interactions over both space (across the mammoth steppe) and time (across the LGM) permit an examination of whether specific patterns of interactions characterized these ecosystems, and to what degree climate change may have influenced the roles of species in these mammalian communities.

We employ a three-pronged approach to address these issues.
First, we use a Bayesian isotope mixing model to quantify the structure, magnitude, and variability of trophic interactions from stable isotope ratios of mammals in six independent Eurasian predator-prey networks spanning the LGM.
Second, we compare species' resource use across the mammoth steppe environment, determine whether these interactions changed in response to the arrival or extinction of co-occurring species, and assess the degree of dietary specialization within and among predator guilds.
Third, we determine how community-level patterns of interaction change from eastern Beringia to Europe across the LGM using recently developed tools from network theory.
In tandem, these combined approaches reveal the variability of mammoth steppe predator-prey network structures, the degree that trophic interactions varied over space and time, and how these changing patterns of interaction influenced the structural properties of mammalian food webs over long timescales.

\section{Materials and Methods} 
\subsection{Study Sites}


During the late Pleistocene, the mammoth steppe extended from western Europe to the Yukon \cite{Bocherens:2003wo}.
The mammalian community that inhabited this steppe has been noted for its species richness \cite{Hopkins:1982va,Guthrie:2001vj} despite the assumed low productivity of local vegetation.
This `productivity paradox' \cite{Hopkins:1982va} suggests that mammoth steppe vegetation differed from modern tundra-dominated flora \cite{Guthrie:2001vj,Yurtsev:2001p3365}.
Indeed, palynological evidence indicates that tundra and boreal vegetation retreated to isolated refugia during the height of the LGM \cite{Guthrie:2001vj,Yurtsev:2001p3365,Brubaker:2005go}. 
It is now generally accepted that before and after the LGM - hereafter the pre-Glacial and post-Glacial, respectively - mammoth steppe vegetation consisted of relatively mesic coniferous woodland mosaics in Beringia and Europe \cite{Brubaker:2005go,Bunbury:2009p3374,Muhs:2001p3359}.
Evidence of forests during the LGM is restricted to south-central Europe \cite{Willis:2004ga}.
In contrast, LGM Beringia was a nearly treeless, hyper-xeric, and highly productive steppe dominated by low-sward herbaceous vegetation \cite{Yurtsev:2001p3365,Zazula:2006hx}.

Mammalian communities were taxonomically similar across Eurasia \cite{Bocherens:2003wo}.
From Beringia to Europe, large felids (the saber-toothed cat, {\it Homotherium serum}, in pre-Glacial Beringia and the cave lion, {\it Panthera spelaea}, in Europe as well as Beringia after the pre-Glacial) \cite{Barnett:2009p3323}, brown bears ({\it Ursus arctos}), and wolves ({\it Canis}) were the dominant predators, while short-faced bears ({\it Arctodus}) were exclusive to Beringia and North America, and cave hyenas ({\it Crocuta crocuta}) were exclusive to western Eurasia and Africa.
Smaller predators, including wolverines ({\it Gulo}) and {\it Lynx} tend to be preserved in European fossil sites.
Mammoth steppe herbivores had similarly large geographic ranges, and included wooly mammoths ({\it Mammuthus primigenius}), caribou ({\it Rangifer tarandus}), yak ({\it Bos mutus}), bison ({\it Bison} spp.), horses ({\it Equus ferus}), caprine bovids ({\it Symbos cavifrons} in Beringia, and {\it Rupicapra rupicapra} in Europe) and the wooly rhinocerous ({\it Coelodonta antiquitatis}).
In contrast to Beringia, Europe hosted a diverse cervid community, including red deer ({\it Cervus elaphus}), roe deer ({\it Capreolus capreolus}), and the Irish elk ({\it Megaloceros giganteus}).
Although the eastern Beringian mammoth steppe ecosystem was not significantly influenced by humans or human ancestors before ca. 13.5 kyr BP \cite{Koch:2006vt}, {\it Homo neanderthalensis} is known to have occupied European systems (sometimes sporadically) from ca. 300 to 30 kyr BP \cite{Mellars:2004cf,Dalen:2012uj}, and modern humans occur in the region at ca. 40 kyr BP \cite{Mellars:2004cf}, including sites on the Arctic Ocean in central Beringia \cite{Pitulko:2004ft}.
Neanderthal diets in continental regions were dependent on terrestrial animals \cite{Bocherens:2001cl,Bocherens:2005jj,Richards:2009p2361}, though their role as predators relative to co-occurring carnivores is not well understood.


\subsection{Estimating diet from stable isotope data}

Ratios of stable isotopes can be used to infer trophic interactions between predators and prey.
Because prey isotope ratios are recorded in consumer tissues, and are robust to diagenetic alteration over long periods of time, they can be used to reconstruct historic or paleontological patterns of resource use \cite{Koch:1998ve,Koch:2007ug,Newsome:2007tz}.
If ratios of stable carbon and nitrogen isotopes are known for both predators and potential prey, and if the fractionation of stable isotopes by metabolic processes between predators and prey is characterized (using trophic discrimination factors), then mixing models can be used to quantify the proportional contribution of prey to a predator's diet \cite{Moore:2008kg}, thereby establishing a per-capita measure of mass flow between interacting species in a food web.
Values of carbon and nitrogen isotope ratios are expressed as ${\rm \delta^{13}C}$ and ${\rm \delta^{15}N}$ respectively, where $\delta = 1000\{(R_{\rm sample}/R_{\rm standard}) - 1\}$ and $R = {}^{13}{\rm C}/{}^{12}{\rm C}$ or ${}^{15}{\rm N}/{}^{14}{\rm N}$, with units of per-mil ($\permil$); reference standards are Vienna PeeDee belemnite for carbon, and atmospheric ${\rm N_2}$ for nitrogen.

We used previously published stable isotope datasets to reconstruct trophic interactions for six independent predator-prey networks from eastern Beringia to western Europe, before, during, and after the LGM (figure \ref{fig_Map}).
The three European predator-prey networks include the Ardennes (ca. 44.7 to 28.7 kyr BP) and Swabian Jura (ca. 44.7 to 28.7 kyr BP) during the pre-Glacial, and Jura during the post-Glacial (ca. 16.9 to 14 kyr BP) \cite{Bocherens:1999ut,Bocherens:2001cl,Bocherens:2011iz}.
Unfortunately, we have no European datasets from the LGM.
The three Beringian predator-prey networks were all located near Fairbanks, Alaska, and date to the pre-Glacial (ca. 50 - 27.6 kyr BP), LGM (ca. 27.6 to 21.4 kyr BP), and post-Glacial (ca. 21.4 to 11.5 kyr BP) \cite{FoxDobbs:2008tq}.
To assess the role of {\it H. neanderthalensis} in pre-Glacial European networks, we used published isotope data for western French and Belgian neanderthal specimens dated to ca. 48 to 34 kyr BP \cite{Bocherens:2005jj,Bocherens:2009wpa,Richards:2009p2361}. 
Because the isotopic values of {\it Equus} and {\it Mammuthus} tissues are similar in the neanderthal sites as well as the Ardennes and Swabian Jura  \cite{Bocherens:1999ut,Bocherens:2001cl,Bocherens:2006un}, we consider an assessment of neanderthal diet from these combined assemblages to be meaningful.
Accordingly, we include neanderthals as potential predators in both pre-Glacial European communities.

\subsection{Dietary analysis}
We used estimates of trophic interactions quantified from stable isotope ratios to reconstruct paleontological predator-prey networks.
Predator-prey networks typically consist of species (nodes) connected by trophic interactions (links).
In quantitative networks, link-strengths describe the relative importance of individual trophic links connecting predators to prey \cite{Yeakel:OfhI8s6v}.
To calculate link-strength distributions for each trophic interaction in a network, we used MixSIR (v. 1.0.4), a Bayesian isotope mixing model \cite{Moore:2008kg}.
In this context, link-strengths represent the proportional flow of biomass from prey to predators, such that the links connecting all prey to a given predator are constrained to sum to one.
Because Bayesian mixing models account for link-strength variance, proportional prey contributions are quantified as posterior probability distributions, thereby accounting for actual ecological variability, variation in trophic discrimination factors, non-unique solutions, and measurement uncertainty \cite{Moore:2008kg}.
Accordingly, each link is described by a unique probability distribution, such that link-strengths have associated probabilities for all interactions in a predator-prey network \cite{Yeakel:OfhI8s6v} (online supplementary materials, appendix I).
We corrected for metabolic fractionations between consumers and prey by applying a range of potential trophic discrimination factors for both Beringian and European systems (online supplementary materials, appendix II).

Herbivores from each paleontological assemblage are assumed to be potential prey for all co-occurring predators.
Although adults of large-bodied taxa such as {\it Mammuthus} and {\it Coelodonta} would escape predation from most consumers, they may represent important scavenged resources, and are included as potential prey for smaller species.
In contrast, cave bears ({\it Ursus spelaeus}) are not considered to be predators, and are included as potential prey for {\it Panthera}, {\it Crocuta}, and {\it H. neanderthalensis} in European systems. 
This distinction is supported by evidence for strong herbivory among cave bears \cite{Bocherens:2006un}, and for predation on cave bears by large-bodied carnivores \cite{Weinstock:1999wq,Diedrich:2009uu}.

To measure the structural organization of predator-prey networks, we quantified the degree of nestedness and modularity for each system.
Nestedness quantifies the extent that specialist predator diets are subsets of generalist predator diets (calculated using the Nestedness based on Overlap and Decreasing Fill metric\cite{AlmeidaNeto:2008tw}).
Nested trophic interactions can arise from groups of predators avoiding prey that fall below different optimal physiological or energetic requirements \cite{MacArthur:1966uda}, due to competitive hierarchies among co-occurring predators \cite{Cohen:1990uda}, and/or as a consequence of body size constraints on predation \cite{Pires:2013ke}.
Modularity, or compartmentalization (calculated as a function of local link density \cite{Zhang:2008vw}; see online supplementary materials, appendix III), is often observed in extant trophic systems \cite{Guimera:2010wb,Thebault:2010jv,Baskerville:2010vn,Yeakel:OfhI8s6v}, and is thought to promote stability \cite{May:1972cp,Thebault:2010jv} by isolating extinction cascades \cite{Stouffer:2011kd}.
To account for link-strengths, measures of nestedness and modularity are evaluated across cutoff values $i$, such that a given property is first measured for the whole network ($i = 0$), and again at successive intervals as weak links are eliminated for higher cutoff values ($i>0$).
Therefore, measurements of structure at high cutoff values correspond to the structure generated by the strongest interacting species in a network \cite{Yeakel:OfhI8s6v}.
This analysis enabled us to examine how network structure is dependent on the strength of trophic interactions in predator-prey networks \cite{Tinker:tpa}.

Many structural properties correlate with network size \cite{Dunne:2002uy}. 
To enable comparisons between networks with variable species richness, we measured relative nestedness and modularity: $\Delta{\mathcal N_i}$ and  $\Delta{\mathcal M_i}$, respectively. 
As before, $i$ refers to the cutoff value, and $\Delta$ measures the difference between the structural measurement of an empirical (isotopic) network and a model network with 
1) the same species richness, and
2) the same predator:prey ratio \cite{Yeakel:OfhI8s6v}.
A value of `0' indicates no difference between the structure of the empirical network and that of the model; if ${\rm \Delta} > 0$, the empirical network has a higher value than expected by chance; if ${\rm \Delta} <  0$, the empirical network has a lower value than expected by chance.

\section{Results} 

To determine the degree that predator-prey interactions varied across space, we first quantified the proportional contribution of prey to predator diets using the stable isotope ratios of predators and prey.
Posterior probability densities that describe the dietary contribution of prey groups to predators present in both Beringia and Europe were compared.
If predator diet was constrained over space, these probability densities were not expected to vary from Beringia to Europe, thus falling on the 1:1 axis when plotted against one-another (figure \ref{fig_Comp}; see online supplementary materials, appendix IV for additional details).

During the pre-Glacial, felids in both Beringia and Europe had relatively low proportional contributions of prey groups that were found in both regions (median contributions for 5 shared taxa, Beringia: 8\%; Europe: 7\%), while ca. 60\% and 65\% of their diets were derived from herbivores unique to each locality, respectively. 
By comparison, both {\it Canis} and {\it Ursus arctos} had posterior probability densities of shared prey that were variable (figure \ref{fig_Comp}A).
We note that the posterior distributions for the presence of Caprine bovids in the diets of {\it Canis} were bimodal in Europe. 
Bimodal link-strength distributions are interpreted as alternative hypotheses of prey contributions, with probabilities given by the densities of prey contribution estimates.

During the post-Glacial, felid prey-contribution distributions revealed strong dependencies on {\it Rangifer} in both locations (median contribution, Beringia: 57\%; Europe: 51\%).
Canids show a strong dependence on {\it Bison} in Beringia, but not in Europe (mean contribution, Beringia: 57\%; Europe: 8\%), whereas post-Glacial ursids were dependent on {\it Rangifer} in Beringia but not in Europe (median contribution, Beringia: 66\%; Europe: 2\%).

The degree of dietary specialization can be used to summarize consumer dietary strategies, and is a useful metric for comparing consumer populations over both space and time \cite{Newsome:WhhVfocb}.
We calculated dietary specialization for predators in each mammoth steppe predator-prey community (Fig \ref{fig_Spec2}).
Dietary specialization ($\epsilon$) ranges from 0, where all prey are consumed in equal proportions (dietary generalist), to 1, where one prey is consumed to the exclusion of all others (dietary specialist; see online supplementary materials, appendix V).
We determined {\it Arctodus} to be a specialist predator in Beringia (particularly in the pre-Glacial; $\epsilon=$ 0.58; this and hereafter are median values), relying primarily on either {\it Rangifer} or {\it Symbos} in the pre-Glacial, and switching to {\it Bos} during the LGM ($\epsilon = 0.35$; electronic supplementary material, figure S1).
{\it Ursus}, by contrast, was a generalist in the pre-Glacial and LGM (pre-Glacial: $\epsilon = 0.22$; LGM: $0.23$), but after the extinction of {\it Arctodus} adopted a more specialized diet on {\it Rangifer} in the post-Glacial ($\epsilon = 0.42$).
{\it Canis} and both Beringian felids had generalist diets spanning the entire time interval ({\it Canis} pre-Glacial: $\epsilon = 0.24$, LGM: $0.23$, post-Glacial: $0.32$; felids pre-Glacial: $\epsilon = 0.22$, LGM: $0.23$; post-Glacial: $0.28$).

In Europe, predators tended to have more specialized diets. 
{\it Canis} (greatest dietary contribution from either {\it Rupicapra} or cervids; $\epsilon = 0.48$) and to a lesser degree {\it Ursus} (greatest dietary contribution from {\it Rupicapra}; $\epsilon = 0.36$) had relatively more specialized diets in the pre-Glacial. 
In the post-Glacial, {\it Gulo} and {\it Lynx} had specialist diets, scavenging (it is assumed) on {\it Mammuthus} ($\epsilon = 0.55$) and specializing on {\it Lepus} ($\epsilon = 0.51$), respectively (electronic supplementary material, figure S2).
The consumption of {\it Lepus} by {\it Lynx} is consistent with observed predator-prey interactions in North America today \cite{Stenseth:1997th}.
In contrast, {\it Crocuta} had variable dietary proclivities in the pre-Glacial (Ardennes: $\epsilon = 0.21$; Swabian Jura: $0.41$), while {\it H. neanderthalensis} had relatively generalist diets (Ardennes: $\epsilon = 0.30$, Swabian Jura: $0.32$, based on a $\delta^{15}{\rm N}$ discrimination factor of $4.5 \permil$, see online supplementary materials, appendix II).
{\it H. neanderthalensis} was primarily consuming {\it Mammuthus} in the Ardennes (46\% median contribution) and both {\it Mammuthus} and {\it Equus} in Swabian Jura (46\% and 26\% median contribution, respectively; electronic supplementary material, figure S3), supporting results reported by Bocherens et al. \cite{Bocherens:2005jj}.
An assessment of Neanderthal diet with a $\delta^{15}{\rm N}$ discrimination factor of $3.5 \permil$ increases estimates of {\it Mammuthus} specialization to 52\% median contribution in the Ardennes, and 73\% median contribution in Swabian Jura.

Both proportional contribution and specialization estimates can be examined for each predator separately, or for the predator guild as a whole, the latter resulting in measurements made across predators in a community.
In Beringia, the across-predator reliance on specific prey showed strong similarities across the entire time interval (figure \ref{fig_Dens}A).
In the pre-Glacial, {\it Bos}, {\it Symbos}, and to a lesser extent {\it Rangifer}, were heavily preyed upon by the predator guild.
After the local extinction of {\it Symbos} during the LGM, {\it Bos} and {\it Rangifer} remained important prey resources, while the proportional contribution of {\it Bison} increased slightly.
Across the interval, {\it Equus} and {\it Mammuthus} had the lowest proportional contribution values.
Dietary specialization of the predator guild as a whole did not change between the pre-Glacial LGM ($\epsilon_g=0.26$ for both, where the subscript $g$ denotes guild; figure \ref{fig_Dens}B). 
Specialization among predators increased in the post-Glacial ($\epsilon_g = 0.35$), indicating a heavier reliance on a smaller subset of prey.
This trend appears to be driven largely by an increase in the importance of {\it Rangifer} to the predator guild (figure \ref{fig_Dens}).
Although European predators did not show consistent trends in prey reliance between the pre- and post-Glacial, predator guild specialization increased in the post-Glacial period, from $\epsilon_g = 0.34$ in both pre-Glacial Ardennes and Swabian Jura, to $\epsilon_g = 0.44$ in post-Glacial Jura (electronic supplementary material, figure S4).

Analysis of relative nestedness ($\Delta {\mathcal N}$) revealed that trophic interactions in Beringian and European predator-prey networks are not more nested than expected by chance (electronic supplementary material, figure S5). 
The absence of nested interactions has been observed in other predator-prey systems \cite{Thebault:2010jv,Yeakel:OfhI8s6v}, and our measurement of nestedness across cutoff values shows that this property is absent in both the whole network (low cutoff values - accounting for both weakly and strongly interacting species) as well as for strongly interacting species (high cutoff values).
Analysis of relative modularity ($\Delta {\mathcal M}$) showed Beringian networks to be modular in the pre-Glacial, particularly for strongly interacting species (where only proportional contributions of prey $\geq$ 0.3 and 0.5 are considered, corresponding to cutoff values 0.3 to 0.5) (figure \ref{fig_Mod}A), non-modular during the LGM, and with some modularity (for cutoff values 0.2-0.3) in the post-Glacial (figure \ref{fig_Mod}A; solid lines).
In this analysis, we consider two $\delta^{15}{\rm N}$ trophic discrimination factors to be equally likely (online supplementary materials, appendix II), however measurements of carnivores in the sub-arctic suggest that this value may be closer to $\Delta^{15}{\rm N} = 4.5\permil$.
If a $\delta^{15}{\rm N}$ TDF closer to $4.5\permil$ is considered for the Beringian systems \cite{FoxDobbs:2008tq}, modularity is increased during both the LGM and post-Glacial (figure \ref{fig_Mod}A; stippled lines).
This suggests that our estimates of modularity may be overly conservative.
In contrast, Europe showed little to no modularity across all cutoff values (figure \ref{fig_Mod}B).

\section{Discussion}

The cooling and drying trends associated with the LGM were particularly significant in northeastern Siberia and Beringia \cite{Guthrie:2001vj}, but had large effects on the environment across the entire mammoth steppe.
Analysis of the organization and magnitude of trophic interactions in mammalian communities before, during, and after the LGM provides insight regarding
1) the extent to which species interactions varied across the mammoth steppe, 
2) if interaction structures, measured on different scales, were impacted by the LGM, and
3) if so, whether these structures returned to a pre-perturbation state after the LGM.
Understanding the flexibility of mammalian predator-prey networks, and whether the interactions that form these systems can be re-established after global climatic perturbations, is relevant to current problems facing modern ecosystems.

\subsection{Spatio-temporal patterns of species interaction}

Our comparison of Beringian and European link-strength distributions show felid diets to be more constrained over space than those of {\it Canis} or {\it Ursus}, particularly in the post-Glacial (figure \ref{fig_Comp}).
{\it Rangifer} became an important component of felid diets in the post-Glacial, coinciding with an observed increase in {\it Rangifer} abundance, particularly in North America ca. 20 kyr BP \cite{Lorenzen:2011gp}, although we cannot rule out that this dietary switch was a consequence of behavioral changes independent of prey population dynamics. 
The strongest dietary estimates, corresponding to the peak densities of prey contribution distributions, for {\it Canis} and {\it Ursus} show different patterns than those of felids, however the increase in {\it Rangifer} abundance may have impacted these predators as well.
The dissimilarity in {\it Canis} and {\it Ursus} diets highlights their ecological plasticity, particularly during the post-Glacial.
Previous studies have shown {\it Canis} to be a generalist predator during the Pleistocene \cite{Leonard:2007iz,FoxDobbs:2008tq}; we show that not only are they generalists at the locality level, but that they are also highly flexible in prey choices in both space and time.
Modern wolves are opportunistic predators \cite{Okarma:1995uwa,Spaulding:1998wsa}, but often specialize on locally abundant cervids \cite{Milakovic:2011hi}.
Some of the variability in Pleistocene canid diets may relate to a greater diversity of wolf ecomorphs.
The eastern Beringian population, for example, had a more robust cranial morphology associated with scavenging \cite{Leonard:2007iz}.
Although the intercontinental ranges shared by felids, {\it Canis}, and {\it Ursus}, are a testament to their success, felids appear to have more constrained diets over the mammoth steppe ecosystem.
If dietary constraints lead to a greater risk of population extinction \cite{McKinney:1997wj,VanValkenburgh:2004p2451,Raia:2012gk}, these differences among taxa may have contributed to the range contraction of large felids across Eurasia, while {\it Canis} and {\it Ursus} retained their spatial distributions into the late Holocene.

While felids consumed a similar diet across the mammoth steppe (especially in the post-Glacial), our quantification of dietary specialization reveals that they were strong dietary generalists, particularly in Europe (figure \ref{fig_Spec2}).
{\it Canis} and {\it Ursus} were also generalist feeders, with some temporal variation. 
In contrast, the short-faced bear {\it Arctodus} was a dietary specialist in the pre-Glacial, relying primarily on {\it Rangifer} (supporting results by Fox-Dobbs et al. \cite{FoxDobbs:2008tq}).
During the LGM, however, {\it Arctodus} prey-contribution estimates reveal a switch towards {\it Bos}, after which the short-faced bear disappears from the fossil record.
It is interesting to note that {\it Arctodus} is the only Beringian predator whose reliance on {\it Rangifer} decreased after the pre-Glacial.
If {\it Rangifer} was a preferred food of {\it Arctodus} (as the pre-Glacial isotope record suggests), a scenario in which short-faced bears were competitively displaced by co-occurring predators is a possibility.

In Europe predator specialization is more variable.
The low {$\epsilon$} values among felids in the pre-Glacial (possibly due to dietary specialization among individuals \cite{Bocherens:2011iz}, which could result in population-level generalization) are similar to those for neanderthals, however prey-contribution results show felids consume relatively greater amounts of {\it Rangifer} (particularly in Swabian Jura), while neanderthals consumed {\it Mammuthus} and {\it Equus}.
We have not considered the impact of {\it Homo sapiens} in European sites, and cannot rule out the possibility that the presence of human hunter-gatherers may have contributed to observed predator specialization.

\subsection{Spatio-temporal patterns of community organization}

Consumption of prey species by the predator guild is strongly consistent in pre-Glacial, LGM, and post-Glacial Beringia (figure \ref{fig_Dens}A).
Although it has been noted that {\it Mammuthus} was under-utilized in Beringia in all time periods \cite{FoxDobbs:2008tq}, our results show a similarly low reliance on {\it Bison} and {\it Equus}.
The low contribution of {\it Bison} may be the consequence of a sharp decline in {\it Bison} abundance beginning ca. 35 kyr BP, and accelerating after 16 kyr BP \cite{Shapiro:2004p2600,Lorenzen:2011gp}.
A shift to a reliance on {\it Rangifer} by the entire predator guild mirrors the dietary switch observed for felids.
There are no consistent patterns of resource acquisition among European predators between pre- and post-Glacial times, but in contrast to the situation in Beringia, {\it Mammuthus} is a more important prey resource across the entire time interval, while {\it Equus} is an important resource in all sites but the Ardennes. 

We find in both Beringia (figure \ref{fig_Dens}B) and Europe that specialization in the predator guild as a whole ($\epsilon_g$) increased in the post-Glacial, indicating that a smaller proportion of available prey species were more heavily used across predator species.
Changes in prey abundance undoubtedly affected predator species differently.
The observed increase in dietary specialization at the guild level indicates a general trend towards increasing resource specialization among predators, coincident with a general decline and range contraction of many Eurasian herbivores \cite{Lorenzen:2011gp}.
In Beringia this trend appears to be largely influenced by the increased contribution of {\it Rangifer} to the diets of predators, while European predators tend to have more idiosyncratic specializations.

\subsection{Linking species interactions to large-scale community structure}

Our analysis of relative nestedness and modularity ($\Delta {\mathcal N}$ and $\Delta {\mathcal M}$, respectively) reveals within-region similarities and between-region differences from the pre- to post-Glacial.
Thus there are large-scale structural differences in the organization of species interactions, despite the presence of similar taxa, across the mammoth steppe.
Nestedness is low in both regions across all time intervals.
As dynamical analyses have shown nestedness to be a destabilizing structure in food webs \cite{Thebault:2010jv,Allesina:2012cv}, its absence may have promoted stability during the dramatic climate changes across the LGM.

Modularity is also low in Europe, but is relatively high in pre-Glacial Beringia, as well as in LGM and post-Glacial Beringia if trophic discrimination factors were large (figure \ref{fig_Mod}).
Modularity in pre-Glacial Beringia chiefly originates from a strong similarity in prey choice by {\it Canis} and {\it Ursus}, whereas felids and {\it Arctodus} have more idiosyncratic diets across the LGM  (electronic supplementary material, figure S1).
Modularity is associated with dynamic stability and increased persistence \cite{Guimera:2010wb,Thebault:2010jv,Stouffer:2011kd}, and implies that the pre-, and perhaps the LGM and post-Glacial Beringian systems were more internally stable than European systems.
Although the modularity of species interactions changes from Beringia to Europe, it does not appear to change much over time, suggesting that the LGM had little impact on the structure of predator-prey networks, despite significant changes in the interactions between species.
It also suggests that mammoth steppe communities were well-adapted to the climatic perturbations associated with the LGM, and is consistent with the notion that climate change was not solely responsible for the end-Pleistocene extinctions \cite{Koch:2006vt,Gill:2009p2368}.

There are two potential explanations that account for the spatial differences in modularity across the mammoth steppe.
In Beringia, the spatial segregation of plant species with either physiognomic differences or preferences for different micro-habitats could result in modular predator-prey interactions.
For example, modern East African food webs are compartmentalized into spatial guilds (woodland vs. grassland) \cite{Baskerville:2010vn}, that are especially pronounced for strongly linked species \cite{Yeakel:OfhI8s6v}.
These spatial guilds have distinct $\delta^{13}{\rm C}$ values because of the differences between $\rm C_3$- and $\rm C_4$-photosynthetic plants.

In Beringia, spatial variability of plant isotope values could arise from
1) isotopic differences in plants inhabiting different micro-habitats (where small differences in humidity, rainfall, or soil moisture may impact the isotopic values of local plant tissues \cite{OLeary:1988p1276}), leading to differences among herbivores that consume plants in these micro-habitats (and by extension, their predators), or
2) isotopic differences among different plant functional types (e.g. shrubs, grasses, lichen), such that the dietary preferences of herbivores result in isotopic differentiation among browsers, grazers, and their respective predators \cite{Bocherens:2003wo,Drucker:2003dz}.
Two lines of evidence suggest that the latter is more likely: 
within-region variation in herbivore dental micro- and mesowear reveal strong dietary differences among herbivores \cite{Rivals:2010p3404}, and
significant differentiation exists in the isotopic values of different plant functional groups \cite{FoxDobbs:2008tq}.
Thus, we conclude that it is more likely that herbivores accumulated distinct isotopic values as a function of dietary differences rather than from foraging in isotopically-distinct micro-habitats.
Although the spatial patterning of vegetation in Beringia is disputed \cite{Guthrie:2001vj,Zazula:2003dna}, there is little support for spatial differentiation of plant functional types at the scale present in African savanna-woodland environments, particularly during the LGM \cite{Guthrie:2001vj}.
This suggests that there may be an alternative explanation for the modularity of Beringian predator-prey networks.

This explanation of modular network structure invokes the insularity of the Beringian mammoth steppe community relative to that of Europe.
Modular food webs are defined by dietary resource segregation among consumers \cite{Guimera:2010wb}.
Resource segregation can occur over ecological, but also evolutionary timescales, where coevolutionary relationships may begin to constrain the plasticity of trophic interactions, promoting compartmentalization \cite{GuimaraesJr:2007wz}.
In isolated environments, where neighboring systems are similar and invasions are rare, differentiation of resources and the subsequent development of modular interactions may be more likely to occur and reinforced over time.
In contrast, systems that are bordered by a diverse array of animal communities and are highly diffuse may be held in a transient state such that niche diversification is continually interrupted, limiting compartmentalization.
We suggest that pre-Glacial - and possibly LGM and post-Glacial - Beringia may have been modular due to stronger homogeneity with, and periodic isolation from, neighboring communities.
This insularity would serve to limit invasions of species from dissimilar communities, allowing consumers to minimize competitive overlap while maximizing resource diversity. 
Europe, by comparison, was an ecological nexus \cite{Agusti:2002ua}, where the periodic influx of species from diverse communities may have limited niche diversification among species, preventing compartmentalization, and resulting in the unstructured predator-prey networks that we observe across the LGM.
We are not aware of any analysis designed to test this specific mechanism for preserving (or disrupting) modularity, and we suggest that this would be a fruitful theoretical exercise.


Modern mammalian communities are remnants of a rich Pleistocene heritage.
Knowledge of the relationships among Pleistocene species will inform our understanding of extant ecosystems.
Moreover, studies of past ecosystems permit an examination of how communities responded to climatic or other external perturbations over long timescales.
Because many species inhabiting mammoth steppe environments are present (and in many cases at risk) in modern ecosystems, reconstruction of the structure of species interactions across the LGM is increasingly relevant for understanding the potential resilience and plasticity of modern species.

\section*{Acknowledgements}
We are grateful to S Allesina, T Gross, AM Kilpatrick, C Kuehn, T Levi, M Mangel, M Novak, M Pires, PI Prado, L Rudolf, AO Shelton, DB Stouffer, and C Wilmers for helpful discussions and/or criticisms. The Advanced Ecological Networks Workshop in S\~ao Pedro, Brazil provided a singular opportunity to discuss many of the ideas that encouraged and improved this work. This research was funded from a National Science Foundation (NSF) Graduate Research Fellowship to JDY, a UC-Santa Cruz Chancellors fellowship to PLK, and the Funda\c{c}\~ao de Amparo \`a Pesquisa do Estado de S\~ao Paulo (FAPESP) to PRG.

\bibliographystyle{plain}
\bibliography{mammoth}

\newpage

\section*{Figure Legends}

\begin{figure}[h!]
   \centering
   \includegraphics[width=1\textwidth]{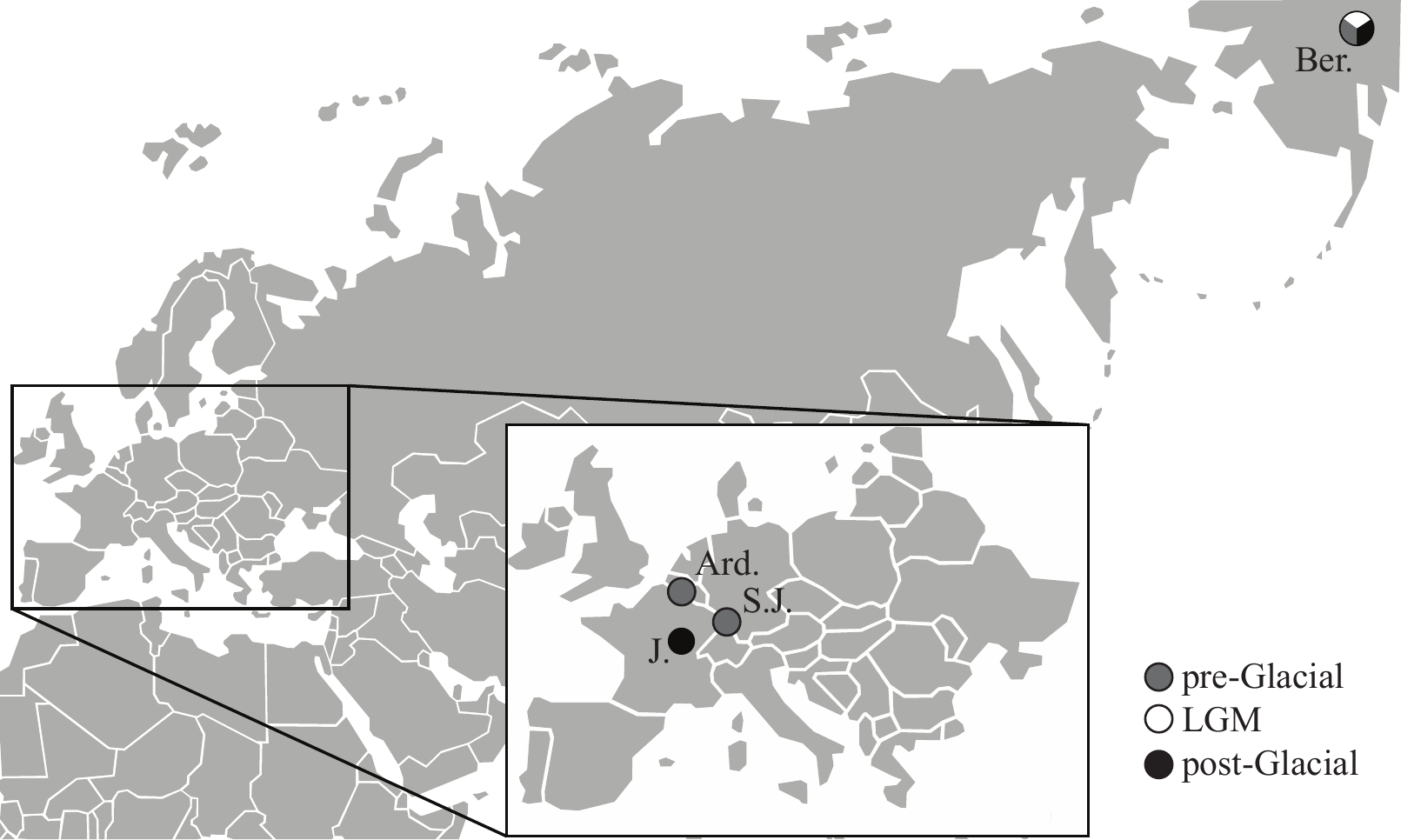}
      \caption{Locations of late Pleistocene mammoth steppe sites included in the analysis. The pre-Glacial, LGM, and post-Glacial Beringian sites are located near Fairbanks, Alaska. Two pre-Glacial and one post-Glacial European site occur in eastern France, Belgium, and western Germany, respectively. Ber. = Beringia; Ard. = Ardennes; S.J. = Swabian Jura; J. = Jura.
      }
      \label{fig_Map}
\end{figure}

\newpage

\begin{figure}[h!]
   \centering
   \includegraphics[width=1\textwidth]{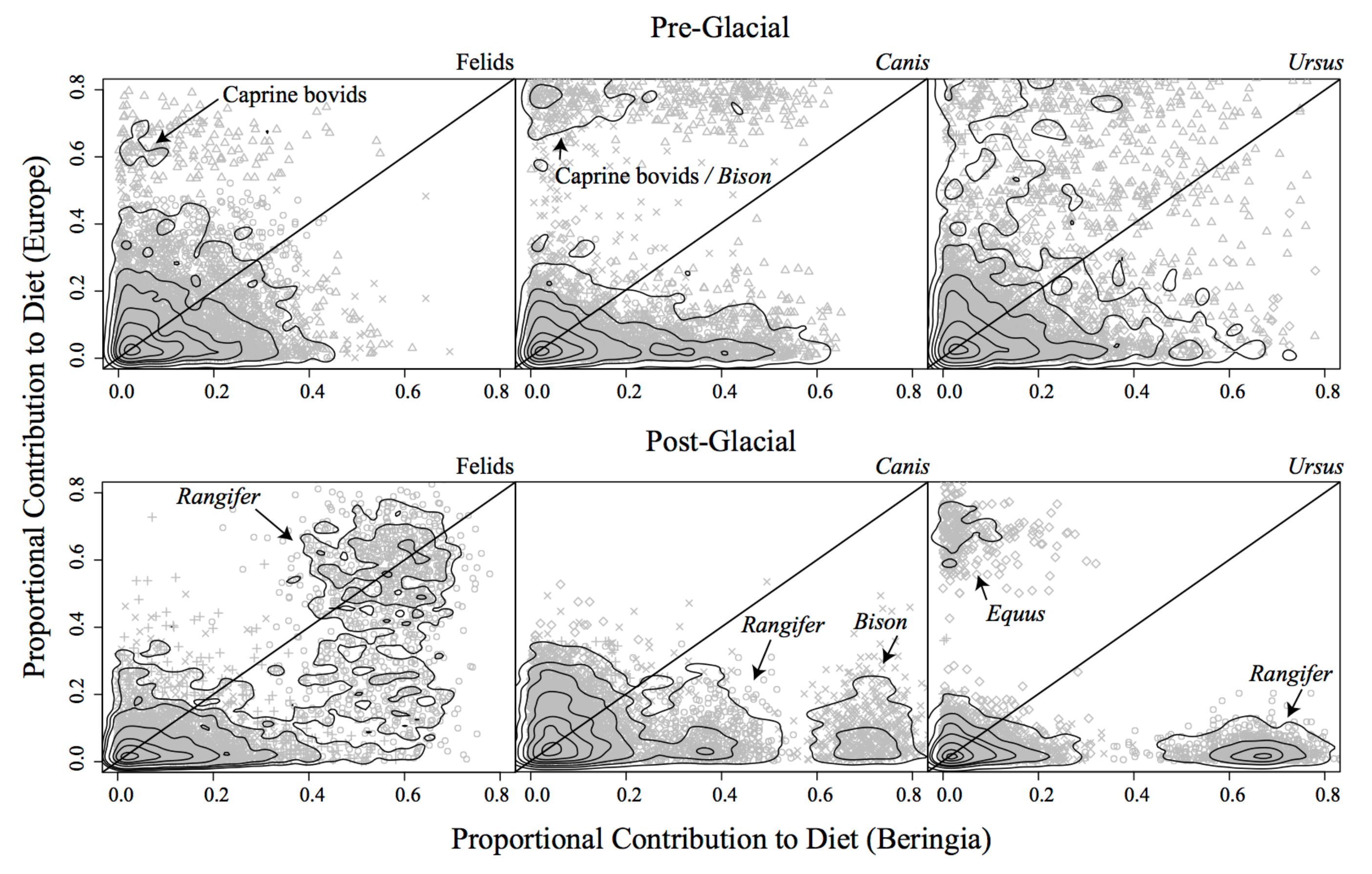}
      \caption{
     Proportional contribution estimates of prey taxonomic groups to the diets of predators present in both Beringia and Europe in the pre- and post-Glacial periods; taxa not present in both localities are not shown. Points represent individual prey-contribution estimates from the Bayesian isotope mixing model, MixSIR ($\diamond$ {\it Equus}, $\times$ {\it Bison}, $+$ {\it Mammuthus}, $\vartriangle$ Caprine bovids, $\circ$ {\it Rangifer}), and contours show the densities of all points. Contributions of prey that do not differ for predators in Beringia and Europe fall on the 1:1 line.
      }
      \label{fig_Comp}
\end{figure}

\newpage

\begin{figure}[h!]
   \centering
   \includegraphics[width=1\textwidth]{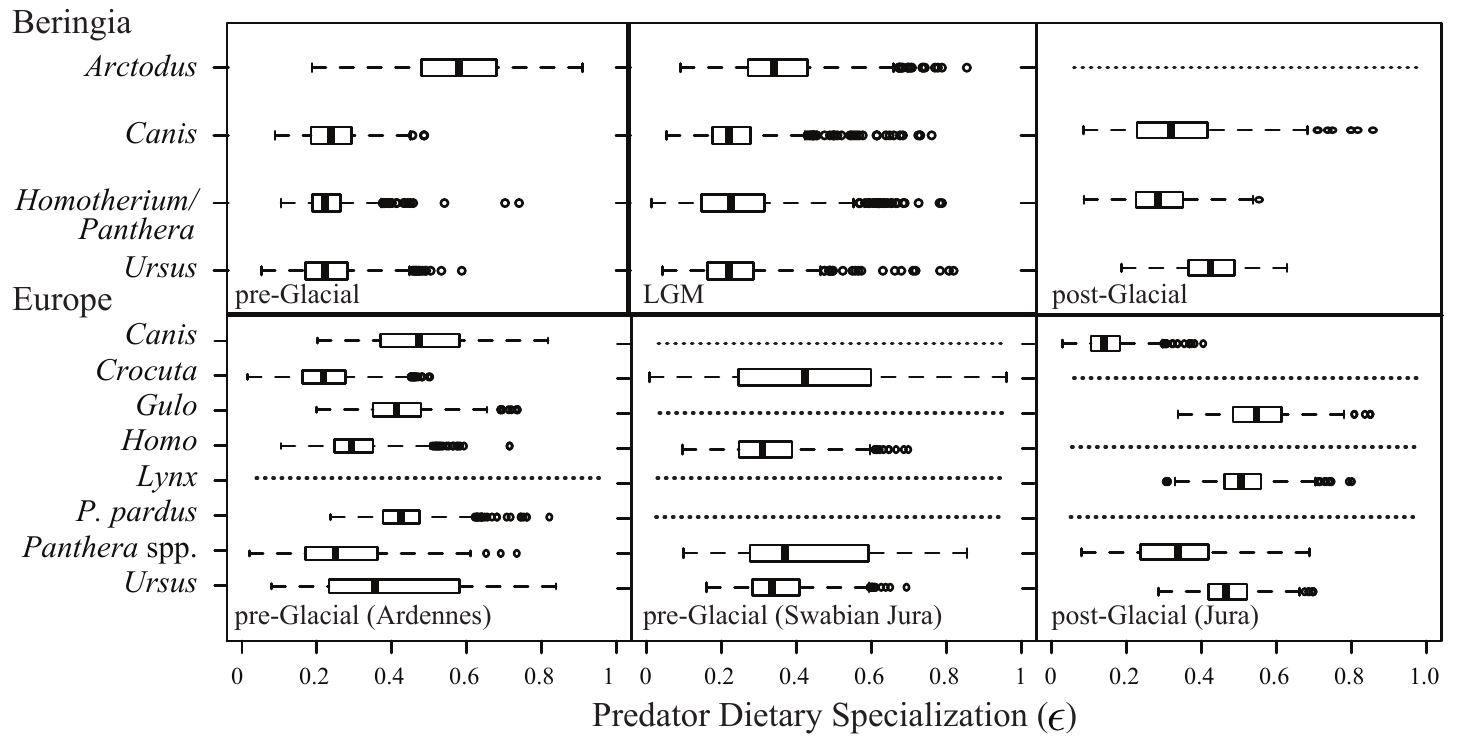}
      \caption{Predator dietary specialization ($\epsilon$) for Beringian and European predator species from the pre-Glacial to the post-Glacial. A value of $\epsilon=0$ describes a generalist diet (consumption of all prey in equal amounts), whereas a value of $\epsilon = 1$ describes a specialist diet (consumption of one prey to the exclusion of others). $\epsilon$ values for {\it Homo} were calculated using a $4.5\permil$ trophic discrimination factor. 
      Dotted lines denote species' absence.
      }
      \label{fig_Spec2}
\end{figure}

\newpage

\begin{figure}[h!]
   \centering
   \includegraphics[width=1\textwidth]{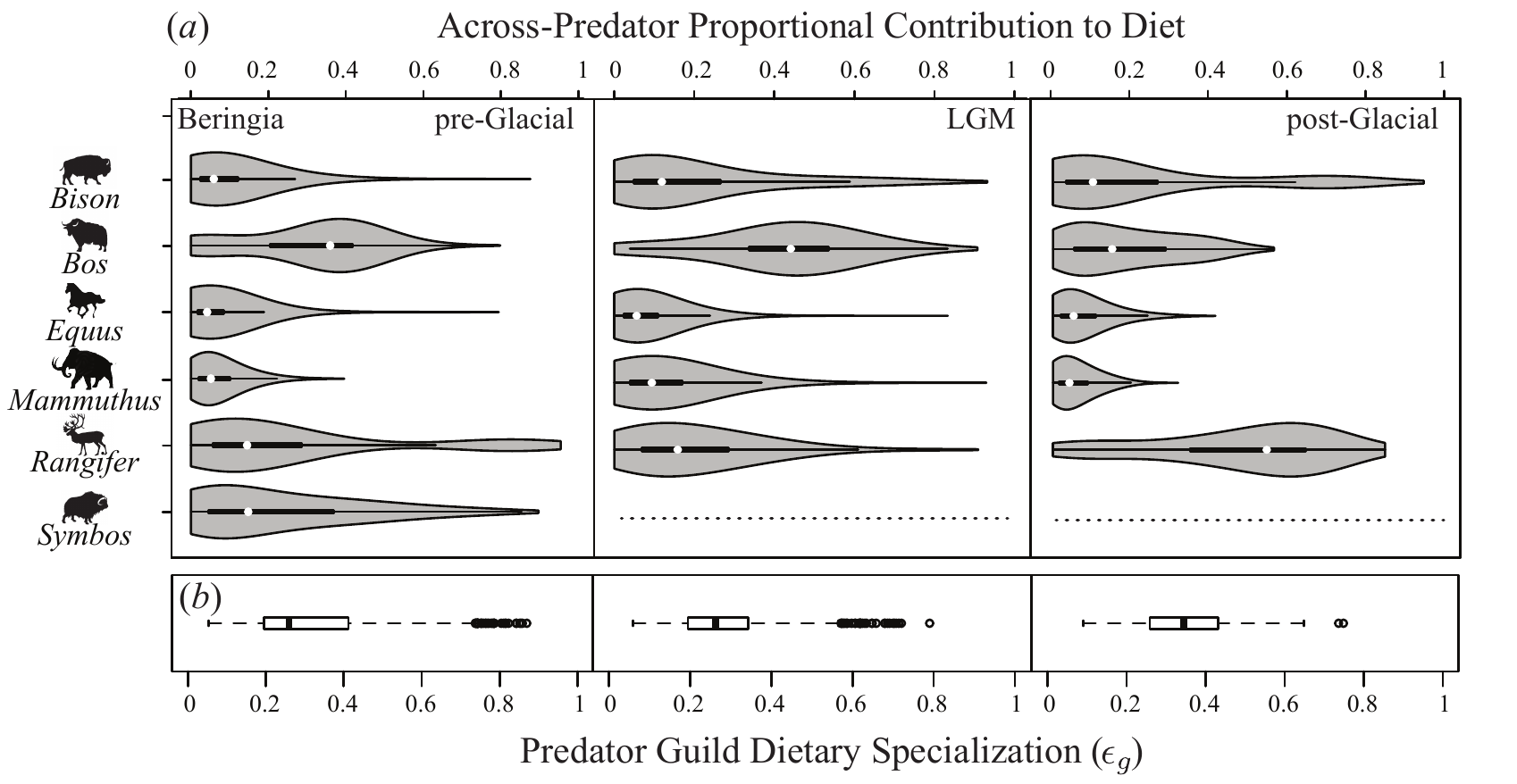}
      \caption{({\it a}) The proportional contribution of each prey across the predator guild in the pre-Glacial, LGM, and post-Glacial Beringia. 
     ({\it b}) Consumer dietary specialization quantified for the predator guild ($\epsilon_g$) for each time period. The median $\epsilon_g$ value is highest for the post-Glacial, indicating an on-average greater contribution of a smaller subset of potential prey; this trend is also observed in European systems (electronic supplementary material, figure S4). Dotted lines denote species' absence.
      }
      \label{fig_Dens}
\end{figure}

\newpage

\begin{figure}[h!]
   \centering
   \includegraphics[width=1\textwidth]{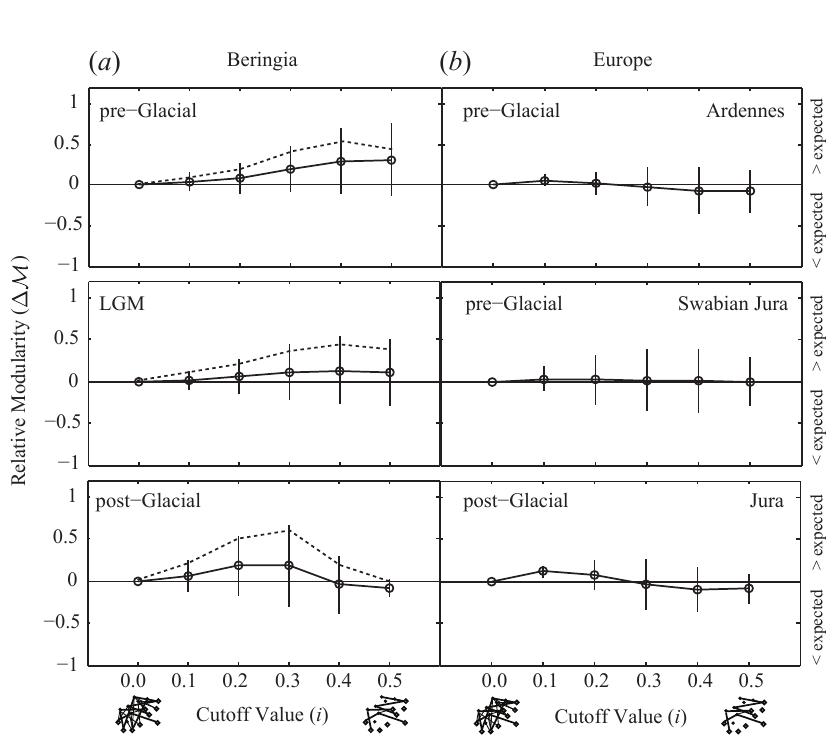}
      \caption{Relative modularity ($\Delta \mathcal{M}$) for ({\it a}) Beringia and ({\it b}) Europe across cutoff values $i$. Values $> 0$ indicate that the system is more modular than is expected given the size and predator:prey ratio of the network; values $< 0$ indicate that the system is less modular than is expected. 
      For Beringian systems, stippled lines denote median relative modularity values if a trophic discrimination factor of $4.5\permil$ is considered. 
      The cutoff value $i = 0$ refers to the whole network with no link deletions. 
      At high cutoff values only the strongest interacting species affect the structure of the network.
      }
      \label{fig_Mod}
\end{figure}

\end{document}